\documentclass[12pt]{article}
\usepackage{graphicx}
\usepackage{float}

\newcommand\pubnumber{NuPhys2015-Vagnoni}

\def\roma{Dipartimento di Matematica e Fisica\\
INFN \& Universit\`a degli Studi Roma Tre, \\
Via della Vasca Navale 84, Rome, ITALY}

\def\Title#1{\begin{center} {\Large #1 } \end{center}}
\def\Author#1{\begin{center}{ \sc #1} \end{center}}
\def\Address#1{\begin{center}{ \it #1} \end{center}}

\newcommand\pubblock{\rightline{\begin{tabular}{l} \pubnumber\\
\end{tabular}}}
\newenvironment{Abstract}{\begin{quotation}  }{\end{quotation}}
\newenvironment{Presented}{\begin{quotation} \begin{center} 
             PRESENTED AT\end{center}\bigskip 
      \begin{center}\begin{large}}{\end{large}\end{center} \end{quotation}}
\def\Acknowledgements{\bigskip  \bigskip \begin{center} \begin{large}
             \bf ACKNOWLEDGEMENTS \end{large}\end{center}}

\def\beq{\begin{equation}}
\def\eeq#1{\label{#1}\end{equation}}
\def\eeqn{\end{equation}}

\def\beqa{\begin{eqnarray}}
\def\eeqa#1{\label{#1}\end{eqnarray}}
\def\eeqan{\end{eqnarray}}

\let\bar=\overbar

\def\Dslash{\not{\hbox{\kern-4pt $D$}}}
\def\dslash{\not{\hbox{\kern-2pt $\del$}}}

\def\msb{{\bar{\ssstyle M \kern -1pt S}}}

\begin{document}
\begin{titlepage}
\pubblock

\vfill
\Title{Neutrino energy reconstruction in long-baseline experiments}
\vfill
\Author{Erica Vagnoni}
\Address{\roma}
\vfill
\begin{Abstract}
Modern experiments aimed at measuring neutrino oscillation
parameters have entered the age of precision. The determination of these
parameters strongly depends on the ability to reconstruct the energy distributions of
the neutrino beams. We compare two different energy reconstruction
techniques: the reconstruction based on the kinematic of the outgoing
lepton and the one based on the calorimetric method. Furthermore, we
analyze realistic detector capabilities, such as energy resolutions,
thresholds and efficiencies, in order to estimate how well they need to be
evaluated to avoid a significant bias in the extraction of the oscillation
parameters.
\end{Abstract}
\vfill
\begin{Presented}

NuPhys2015, Prospects in Neutrino Physics

Barbican Centre, London, UK,  December 16--18, 2015

\end{Presented}

\vfill
\end{titlepage}
\def\thefootnote{\fnsymbol{footnote}}
\setcounter{footnote}{0}

\section{Introduction}
Neutrino scattering off a nucleus in charged current (CC) leads to the production of the associate lepton and a hadronic final state of $n$ nucleons knocked out from the nucleus and $m$ mesons produced in the process.\\
Neutrino energy can be reconstructed using the lepton kinematic with the assumption that the invariant hadronic mass $W^2$ is known. Thus, applying energy and momentum conservation  
\begin{equation}
\label{eq:kin_rec}
E_\nu^{{kin}} = \frac{W^2 - m^2_\ell + 2(nM- \epsilon_n)E_\ell - (nM-\epsilon_n)^2}{2 (nM -\epsilon_n  - E_\ell +  |\mathbf{k}_\ell | \cos \theta_\ell )} \ ,
\end{equation}
where $E_\ell$ is the energy of the outgoing lepton, $\mathbf{k_\ell}$ its momentum and $\theta_\ell$ its angle with respect to the direction of the incoming beam. $\epsilon_n$ represents the average single-nucleon separation energy. 
A deeper knowledge of the final state, and of the deposited kinetic energies by the particles, could allow a more accurate  reconstruction of the neutrino energy
\begin{equation}
\label{eq:cal_rec}
E_\nu^{cal} = \epsilon_n + E_\ell + \sum_{i}^N (E_{{\mathbf{p_i}'}} - M) +\sum_{j}^M E_{\mathbf{h_j}'} \ ,
\end{equation}
where $E_{{\mathbf{p_i}'}}$ and $E_{\mathbf{h_j}'}$ denote the energies of the \textit{i}th knocked-out nucleon  and of the \textit{j}th produced meson, respectively. \\
The two reconstruction schemes in Eqs. (\ref{eq:kin_rec}) and (\ref{eq:cal_rec}) are employed to analyze the events generated with the Monte Carlo event generator GENIE 2.8.0 + $\nu T$ \cite{genie1} \cite{genie2}. Thus, it is possible to produce migration matrices, $\mathcal{M}_{ij}$, that define the probability for an event with true energy in the \textit{j}th bin to be reconstructed in the \textit{i}th energy bin.\\
Neutrino events are reconstructed assuming: a {\textit{Perfect Scenario}} in which all the particles produced by the interaction are detected, and a {\textit{Realistic Scenario}} where realistic detector capabilities are used for the analysis of the final state.
\section{Oscillation Analysis}
\begin{figure}[H]
\centering
\includegraphics[height=2.9in]{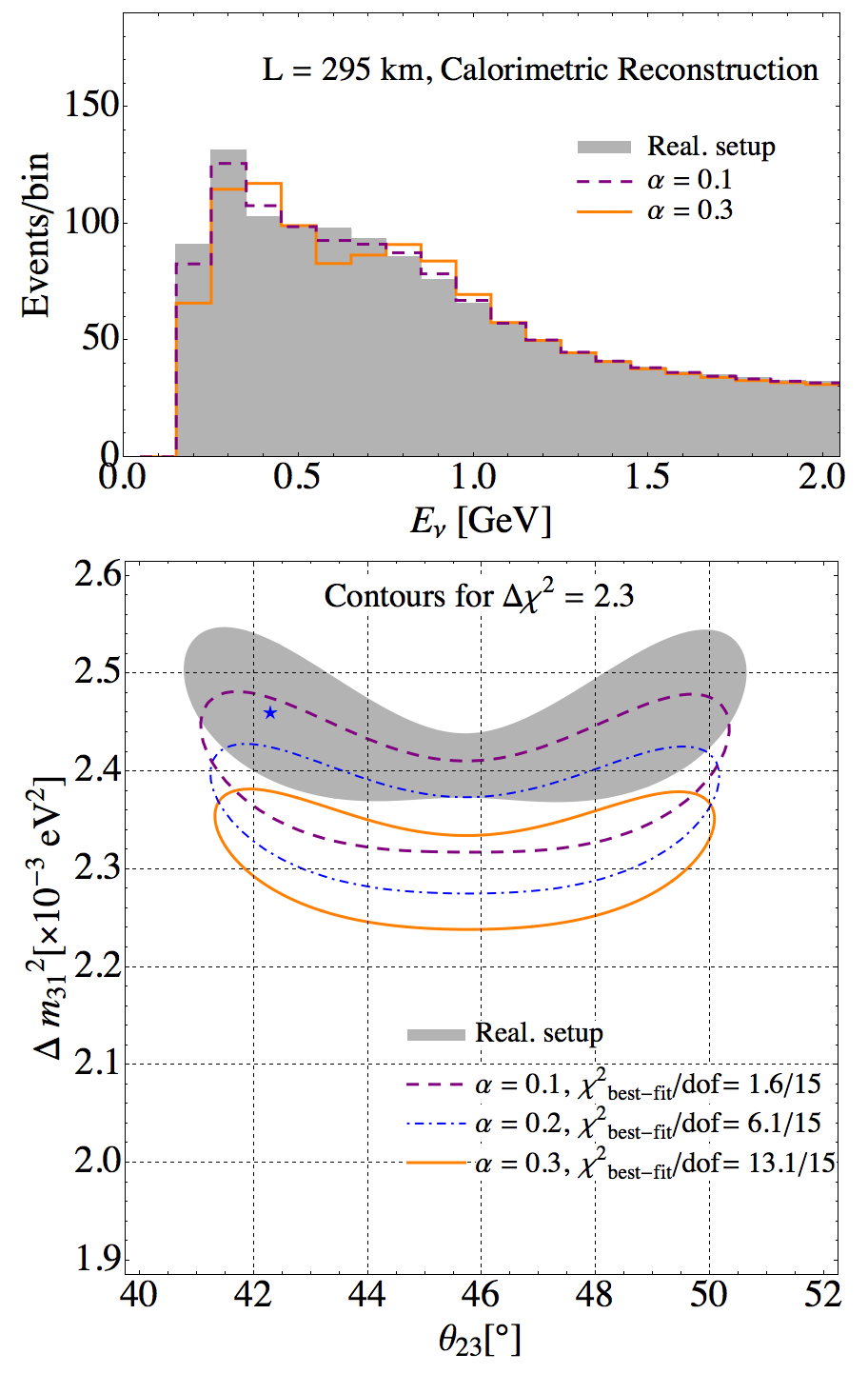} 
\includegraphics[height=2.9in]{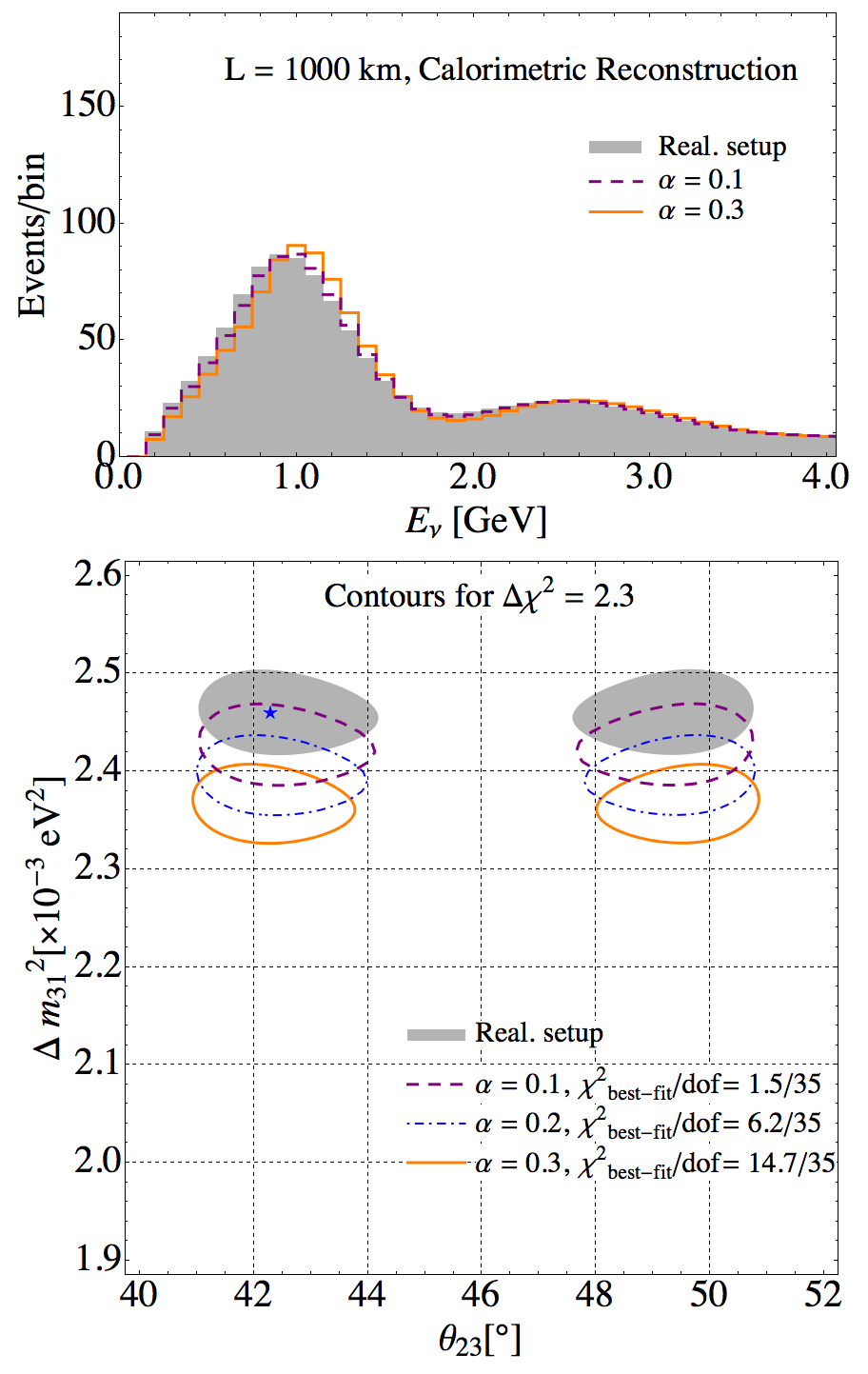}\\
\includegraphics[height=2.9in]{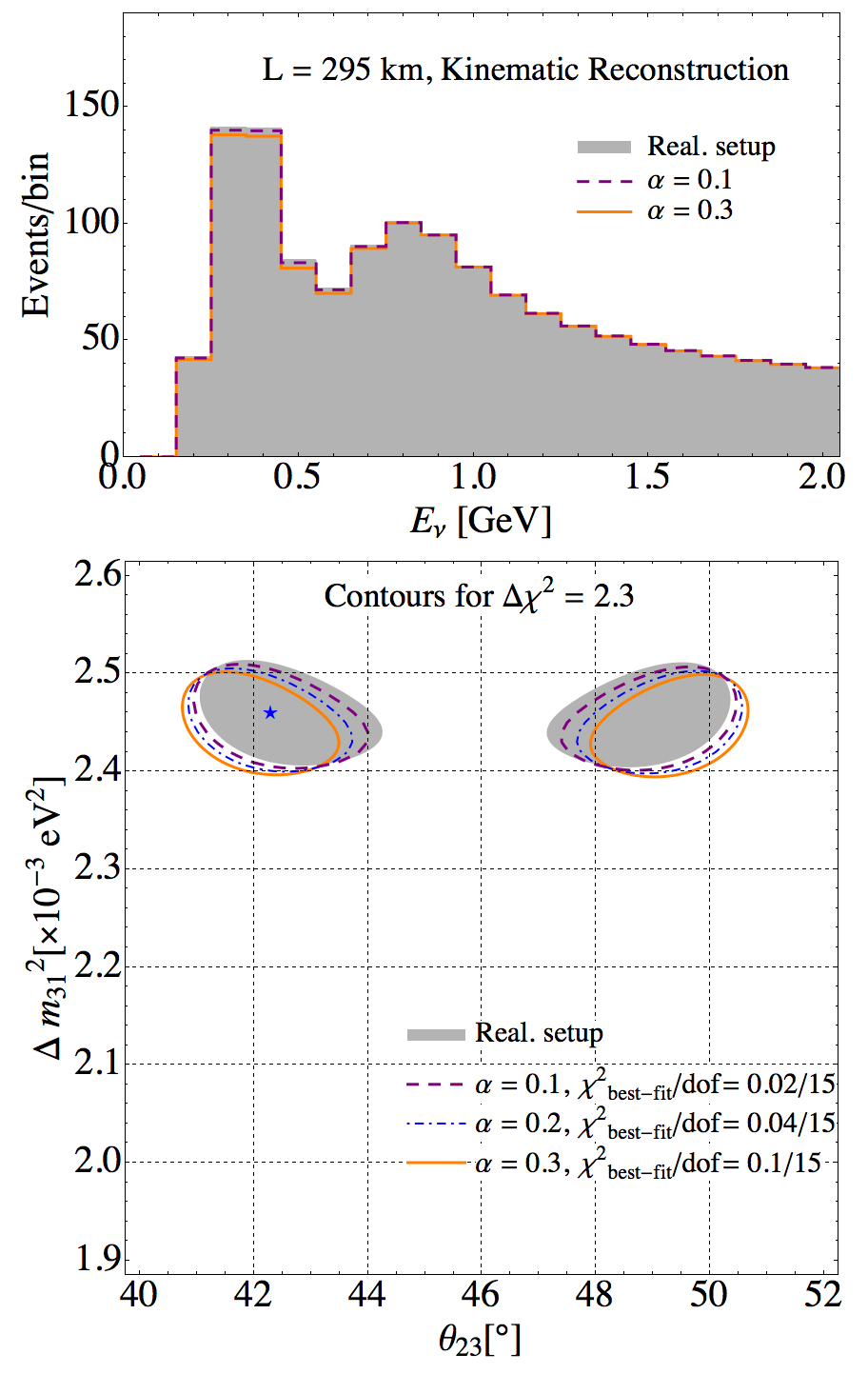} 
\includegraphics[height=2.9in]{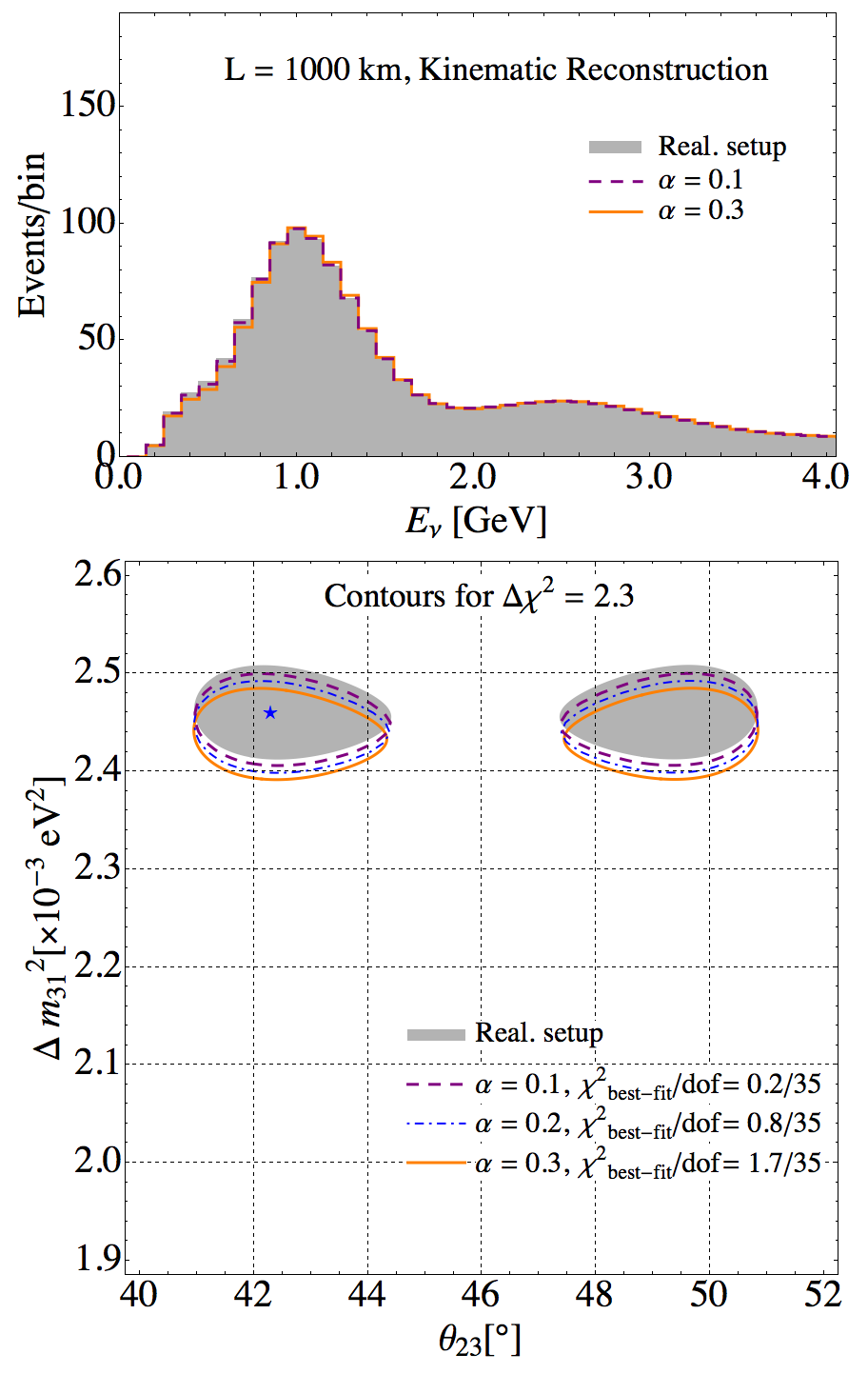}
\caption{Results for the calorimetric and kinematic reconstruction for the two experimental setup. Upper panels: expected event distributions at the far detector. The shaded histograms represent the rates obtained applying migration matrices from the realistic scenario. The dashed and solid lines represent the rates obtained for an underestimation of the detector effects of the $10\%$ and $30\%$, respectively. Lower panels: confidence regions at 1 $\sigma$ with $\Delta \chi^2 = 2.3$, in the $(\theta_{23}, \Delta m_{31}^2)$ plane. The shaded areas are obtained when detector effects are fully estimated. The closed lines are obtained when the fit is performed using a distribution obtained through the combination of perfect and realistic matrices. They represent the contour for detector performances overestimated of the $10\%$, $20\%$ and $30\%$.}
\label{fig1}
\end{figure}

\begin{figure}[H]
\centering
\includegraphics[height=2.3in]{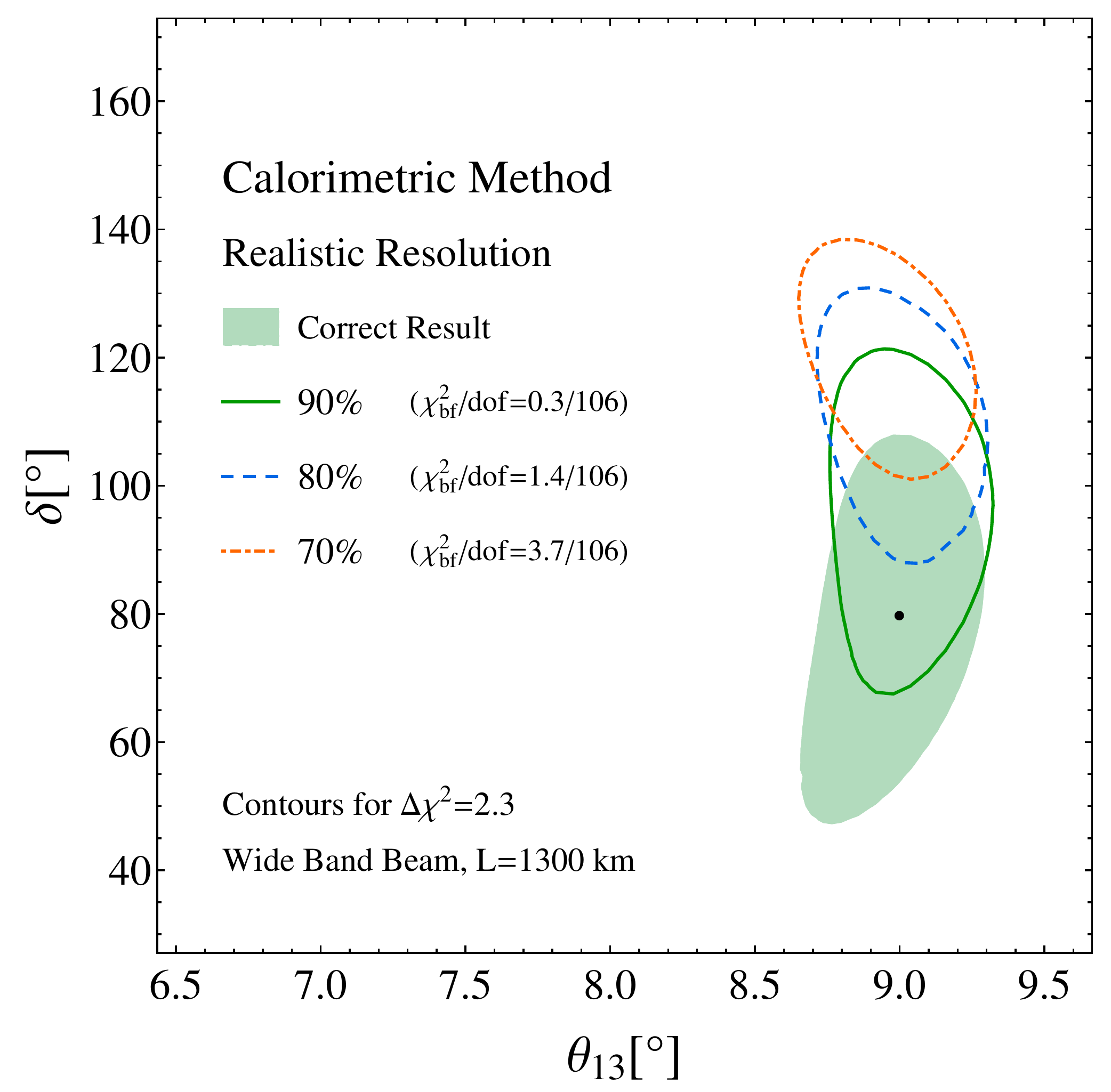}
\caption{Confidence regions in the plane $(\theta_{13},\delta)$ at 1 $\sigma$ for $\Delta \chi^2 = 2.3$. The shaded area is obtained when the missing energy is fully estimated. The lines represent the confidence regions when only the $90\%$, $80\%$ and $70\%$ of the missing energy is correctly accounted for.}
\label{fig2}
\end{figure}

The analysis of the oscillation parameters is performed using the software GLoBES \cite{GLB1} \cite{GLB2}, in the oscillation channel $\nu_\mu \to \nu_\mu$ \cite{paper1}.
The assumed true values of the oscillation parameters used for the analysis are taken from \cite{nufit}.
Two different experimental configurations are considered \cite{LE} \cite{HE}, and their main features are reported in Tab. \ref{setup}.
\begin{table}[t]
\begin{center}
\begin{tabular}{c c  c c c}
 \textbf{Experimental setup} & \textbf{Type} & \textbf{Baseline} & \textbf{Energy Peak} \\
\hline
{Low Energy}  & off-axis  & L = 295 km & 600 MeV \\
{High Energy} & on-axis & L = 1000 km & 1-2 GeV \\
\hline
\end{tabular}
\caption{Details of the two experimental configurations used to perform the oscillation analysis.}
\label{setup}
\end{center}
\end{table}
The number of un-oscillated CC $\nu_\mu$ events for the two different setup used is $\sim 5000$.
The \textit{true} event rates are computed using migration matrices generated within the realistic scenario, with the aim of reproducing  a "realistic" experimental setup  
\begin{equation}
\nonumber
{N_i^{true} = \sum_Y \sum_j \mathcal{M}_{ij}^{Y, real} N_j^Y} \ ,
\end{equation}
for each interaction channel $Y$ considered: quasi-elastic, $2p-2h$, resonance production and deep ineslatic scattering.   
The \textit{fitted} rates are then generated using a linear combination of the matrices obtained from the realistic and perfect reconstruction, as a function of the parameter $\alpha$
\begin{equation}
\nonumber
{N_i^{fit} =  \sum_Y\sum_j \{ (1- \alpha)\mathcal{M}_{ij}^{Y, real} + \alpha \mathcal{M}_{ij}^{Y, perf} \} N_j^Y} \ .
\end{equation} 
This phenomenological approach  is useful to quantify the impact of the incorrect estimation of detectors effect on the oscillation parameters analysis. The obtained results are shown in Fig. \ref{fig1}, for calorimetric and kinematic reconstructions.\\
A similar analysis, has been performed in the appearance channel $\nu_\mu \to \nu_e$ \cite{paper2}. The experimental setup chosen is a wide band neutrino beam with a baseline of $L = 1300$ km, and the neutrino energy is reconstructed applying the calorimetric method. The true event rate is obtained with realistic migration matrices, as before.\\
In the ideal case in which all the particles in the final state are detected, the neutrino energy will be smeared according to a gaussian distribution with a width dependent on the energy smearing assumed for the final state's particles. To estimate missing energy effects, the distribution obtained for the fit are obtained as a linear combination of realistic matrices and gaussian distributions, centered around the true neutrino energy. The result in the $(\theta_{13},\delta)$ plane is shown in Fig. \ref{fig2}.
\section{Conclusions}
We find that the kinematic reconstruction is much robust with respect to detector effects, mostly because muons are well reconstructed. On the other hand, the calorimetric reconstruction strongly depends on the assumed detector performances and its uncertainties can considerably affect the extracted oscillation parameters.\\
We also studied the effects of the missing energy in the appearance channel. To avoid an appreciable bias in the extraction of $\delta$, the missing energy should be correctly estimated at the $90\%$, and if just a $70\%$ is correctly accounted for, the true value would be excluded between 2--3 $\sigma$.

\Acknowledgements
This work has been carried out in collaboration with A. M. Ankowski, O. Benhar, P. Coloma, P. Huber, C.-M. Jen, C. Mariani and D. Meloni.

\end{document}